%Paper: hep-th/9410091
%From: Dmitry Ivanov <ivanov@landau.ac.ru>
%Date: Thu, 13 Oct 1994 11:25:47 +0300
%Date (revised): Thu, 29 Dec 1994 20:22:03 GMT

%
% This is plain TeX file.
%

\def\G {{\cal G }}
\def\DL {{\cal L }}
\def\DR {{\cal R }}
\def\Res#1{\hbox{Res}_{#1}}
\rightline{hep-th/9410091}
\bigskip
\centerline{\bf Knizhnik-Zamolodchikov-Bernard equations on Riemann surfaces}
\bigskip
\centerline{D.Ivanov}
\centerline{\it Landau Institute for Theoretical Physics,
     Moscow 117334, Russia}
\centerline{\it and Institute for Theoretical and
     Experimental Physics, Moscow 117259, Russia}
\centerline{e-mail: ivanov@landau.ac.ru}
\bigskip
\centerline{October 12, 1994.}
\bigskip
{\narrower\smallskip\noindent
Knizhnik-Zamolodchikov-Bernard equations for twisted conformal
blocks on compact Riemann surfaces with marked points are
written explicitly in a general projective structure in terms
of correlation functions in the theory of twisted b-c systems.
It is checked that on the moduli space the equations
provide a flat connection with the spectral parameter.
\smallskip}

\bigskip
\centerline{1. Introduction.}
\bigskip
\nobreak
In addition to the conformal symmetry, the Wess-Zumino-Novikov-Witten
(WZNW) model possesses the symmetry
of an affine Lie algebra $\hat\G$ [1]. The Virasoro algebra is embedded in
$U(\hat\G)$ by the Sugawara construction. This additional symmetry
leads to certain equations for the conformal blocks in the WZNW theory.
When the theory is defined on the sphere, these are the well-known
Knizhnik-Zamolodchikov (KZ) equations [2]:
$$\biggl({\partial\over\partial z_\alpha}+{1\over k + h^*}
\sum_{\beta\ne\alpha} {t_\alpha^a t_\beta^a \over z_\alpha-z_\beta}\biggr)
\langle\Phi_1(z_1)\dots \Phi_n(z_n)\rangle =0.
\eqno (1.1)$$
These equations relate the dependence of the conformal block
$\langle\Phi_1(z_1)\dots \Phi_n(z_n)\rangle$ on the positions
of the marked points $z_\alpha$ to the action of the currents $j^a(z)$ on
the fields $\Phi_\alpha(z_\alpha)$:
$$ t^a \Phi(w)=\oint_w j^a(z)\Phi(w){dz\over 2\pi i}.
\eqno (1.2)$$
We may also speak of the KZ equations as of the connection
$$ \nabla_\alpha={\partial\over\partial z_\alpha}+{1\over k + h^*}
\sum_{\beta\ne\alpha} {t_\alpha^a t_\beta^a \over z_\alpha-z_\beta}
\eqno (1.3)$$
on the bundle of conformal blocks over the moduli space
of the sphere with marked points.

The subject of this paper is the generalization of the KZ equations (1.1)
to surfaces of higher genera and the discussion of their properties.
First these equations for the surfaces of nonzero genera were obtained by
Bernard [3,4]. To define the action of the zero modes of the current
he included twists on the handles of the surface. Conformal blocks become
functions of the twists, and their derivatives along the twists give
the action of the zero modes of the current. The moduli space contains
not only the positions of the marked points, but also the moduli of the
complex structure of the surface itself. The connection form becomes
a differential operator with respect to the twists. We shall call
such a system of the equations the Knizhnik-Zamolodchikov-Bernard (KZB)
equations, and the corresponding connection --- the KZB connection.

In the paper we write out the KZB equations in a simple form. All terms
in the equations can be expressed explicitly as Poincar\'e series
in the Schottky parametrization, and at the same time admit an
invariant description in an arbitrary projective structure. We also
briefly discuss the transformation properties of the equations as
the projective structure changes. A remarkable feature of the KZB
equations is their relationship to the twisted b-c system of spin 1.
After multiplying the conformal block
by the square root of the b-c holomorphic partition function, the
connection depends on the level $k$ only through a spectral parameter
$(k+h^*)^{-1}$; besides the connection form becomes symmetric. It was
pointed out by Losev that this relationship to the b-c systems
can be explained in terms of the BRST construction in the $G/G$
coset model [5]. The presence of the spectral parameter in the connection
ensures that from its flatness at integer $k$ (for ``physical''
reasons) the flatness at an arbitrary $k$ will follow [5]. This imposes
certain strong conditions on the connection form. We explicitly check
these conditions and prove the flatness.

At higher genera the KZB equations contain an interesting ``potential''
term. This term vanishes  at genus one (on a torus), and Bernard
originally claimed that it was zero at any genus [4]. However there
exist indications that the potential should not vanish at higher
genera [5]. So far we cannot say much about the potential term,
except its closeness as a 1-form on the moduli space, which is
one of the conditions for the compatibility of the KZB equations
(cf. [8]).

The paper is organized as follows.
In section 2 we introduce the twisted WZNW model on a compact
Riemann surface.
In section 3 we recall how the stress-energy tensor defines the connection
on the moduli space and how this connection depends on the choice of the
projective structure.
Section 4 deals with twisted 1-forms on the surface; we also define the
twisted b-c system of spin 1 in this section.
These sections provide a necessary kit
for working with the KZB equations. Here we tried to follow
the ideas and the notation of Bernard [3,4] and Losev [5].
Finally, section 5 contains the main results of the paper.
In this section we write out the equations and discuss their properties.
Section 6 summarizes the discussion and presents
several questions for further investigation.
Some auxiliary information and calculations are gathered in Appendices.
The Schottky parametrization of Riemann surfaces is reviewed in
Appendix A.
In Appendix B we derive the KZB equations.
Appendix C contains the proof that the potential
term is a closed 1-form on the moduli space.
In Appendix D we commute the derivatives with respect to the positions
of marked points.
In Appendix E we check the compatibility of the equations for the
partition function which yields the proof of the compatibility
in the general case.

\bigskip
\centerline {2. Twisted WZNW theory.}
\bigskip
\nobreak
We consider the WZNW model at level $k$ on a compact Riemann surface
of genus $N$. The model is defined for a simple compact Lie group $G$ with
the Lie algebra $\G$ [2].
We shall deal only with the holomorphic part of the theory,
thus all fields are holomorphic except at locations of other fields.
The holomorphic theory
contains the currents $j(z)$ taking values in the Lie algebra $\G$,
the stress-energy tensor $T(z)$, and primary fields $\Phi(z)$.
The primary fields
take values in irreducible finite dimensional representations of $\G$
and are multivalued on the surface. The fields obey the operator
product expansion (OPE):
$$ j^a(z) j^b(w) = -k{\delta^{ab} \over (z-w)^2} -{f^{abc}\over
z-w} j^c(w) + O(1),
\eqno (2.1) $$
$$ j^a(z)\Phi(w)= {1\over z-w}t^a \Phi(w) +O(1),
\eqno (2.2) $$
where $\delta^{ab}$ is the invariant bilinear form on $\G$, $t^a$ is
the action of $a\in \G$ in the representation of the primary field $\Phi$,
$f^{abc}$ are the structure constants of $\G$. We shall work in an
orthonormal basis of $\G$ normalized so that
$$ f^{abc} f^{abd} =2h^*\delta^{cd},
\eqno (2.3) $$
where $h^*$ is the dual Coxeter number of $\G$.
The stress-energy tensor $T(z)$ is expressed in terms of the currents by
the Sugawara construction:
$$ T(z)=-{1\over 2(k+h^*)}:\! j^a(z) j^a(z)\! :\,=-{1\over 2(k+h^*)}
\lim_{w\to z} \left(j^a(z) j^a(w) + {k \dim\G \over (z-w)^2}\right).
\eqno (2.4) $$

To be specific about notation, we shall label normalized correlation
functions (divided by the partition function) with the subscript $N$,
like $\langle X \rangle_N$, non-normalized correlators being
without any subscript: $\langle X\rangle=\langle X\rangle_N Z$, where
$Z=\langle 1\rangle$ is the partition function.

Our ultimate goal is to study the connection on the bundle of correlation
functions over the moduli space. To write this connection is
the same as to compute the
correlation functions with the insertion of $T(z)$ (see section 3).
Since $T(z)$ is constructed of the currents, to write the
KZB connection we first need to compute correlation functions
with the currents inserted. On the sphere any correlation function
$\langle j^a(z)X\rangle$ is a 1-form in $z$ which can be
reconstructed from its singularities (singular terms in Laurent
expansion at the poles). On Riemann surfaces of
higher genera there exist global holomorphic 1-forms, and to restore
a meromorphic 1-form, we need extra information, besides singularities.
For example, any meromorphic 1-form is uniquely defined by
its singularities and integrals over A-cycles. Here we encounter
difficulties, since zero modes
$$\oint_{A_i}\langle j^a(z)X\rangle \, dz
\eqno (2.5) $$
are not determined by the OPE.
For this reason Bernard suggested to include twists on
A-cycles [3,4]. Given a set of $N$ elements of $G$
($N$ is the genus of the surface)
$$ g_i=\exp\Bigl(\sum_a \xi_i^a t^a \Bigr), \qquad i=1,\dots,N,
\eqno (2.6) $$
we insert in all correlation functions the twists
$$ \hat g_i = \exp \oint_{A_i} \xi_i^a j^a(z) \, dz.
\eqno (2.7)$$
{}From now on we shall suppress the twists in notation, thus the correlators
$\langle X \prod_{i=1}^N \hat g_i\rangle $ will be written simply as
$\langle X \rangle$.

Following Bernard, we write equations not for a single
partition function or a correlation function, but for them
as functions of the twists. This allows us to express zero modes
of the currents as the derivatives along the twists:
$$\langle\oint_{A_i} dz \, j^a(z) X\rangle = \DL^{ia}\langle X\rangle,
\eqno (2.8) $$
where $\DL^{ia}$ is the right-invariant derivative along the $i$-th twist:
$$\DL^{ia} f(g_1,\dots,g_N)={\partial \over\partial \xi}\Big\vert_{\xi=0}
f(g_1,\dots,e^{\xi t^a} g_i,\dots,g_N)
\eqno (2.9)$$
for any function $f$ of $N$ group elements $g_i$.

\bigskip
\centerline{3. 1-forms on the moduli space and projective structures}
\centerline{on the surface.}
\bigskip
\nobreak
The variation of a correlation function under an infinitesimal shift of
the moduli is described by inserting the stress-energy tensor
$T(z)$. Indeed, let us consider an infinitesimal change of
complex structure induced by a coordinate transformation
$$ z \mapsto z +\varepsilon(z,\bar z),
\eqno (3.1) $$
where $\varepsilon(z,\bar z)$ is defined only locally
(otherwise it would give a reparametrization). The Beltrami
differential
$$ \mu(z,\bar z) =\bar\partial\varepsilon(z,\bar z)
\eqno (3.2) $$
is defined globally and has the transformation properties of a
(-1,1)-form on the surface. At the same time it may
be thought of as a tangent vector to the moduli space.
To compute the variation of a (non-normalized) correlation
function under the change of moduli described by $\mu$,
we need to insert the stress-energy tensor coupled to
$\mu$ inside the correlator:
$$\delta_\mu \langle X \rangle =\int_\Sigma \langle T(z)X \rangle
\, \mu(z,\bar z) \, dz d\bar z.
\eqno (3.3) $$
This expression would be well-defined if $T(z)$ were a 2-differential on
the surface $\Sigma$. In fact, the holomorphic
stress-energy tensor $T(z)$ is not a 2-differential, but transforms with
the Schwarzian term:
$$ T(w)=\left( {dz\over dw} \right)^2 T(z) +
{c \over 12} \{z;w\},
\eqno(3.4) $$
where
$$\{z;w\}=\left( {d^3 z\over dw^3}\Big/{dz\over dw} \right)
-{3\over 2} \left( {d^2 z\over dw^2}\Big/{dz\over dw} \right)^2,
\eqno(3.5) $$
$c$ is the Virasoro central charge. The Schwarzian derivative
$\{z;w\}$ vanishes for projective transformations, therefore
after fixing the projective structure $T(z)$ becomes a 2-differential.
Thus the values of the correlation function
depend on the choice of the family of projective structures on the
surfaces [7]. Physically, this is due to the anomaly;
prescriptionally, this dependence appears in the regularization of
the stress-energy tensors (2.4), (4.18). One easily checks that
the normal ordering in (2.4) and (4.18) depends on the choice
of the local coordinate, and that this dependence reproduces
exactly the transformation law (3.4).

The projective structures should be chosen in such a way that the system
of equations (3.3) is compatible. Then the KZB equations would be
compatible, i.e. they would define a flat connection (not just
projectively flat). Such families of projective structures exist.
One of them is the projective structure defined by the
Schottky parametrization
(see Appendix A). The Schottky parametrization is also convenient
for writing out explicit formulas, and we shall present the expressions
for all terms of the equations in the Schottky representation.
However, we must stress that the whole treatment is parametrization
independent, and can be performed in any ``compatible'' projective
structure (i.e. such that the equations (3.3) are integrable). Any
change of the projective structure is defined by adding a 1-form
on the moduli space to $T(z)$, therefore one can easily construct
a ``non-compatible'' projective structure by adding a non-closed
1-form. From now on we fix a ``compatible'' projective structure
and speak of $T(z)$ as of a 2-differential, and of correlation
functions as of functions on the moduli space.

Remark now that $T(z)$ being a 2-differential on the surface serves as a
1-form on the moduli space [7]. Indeed, by (3.3) it defines the
response to {\it any} change of the moduli. In particular, if we
change the moduli by moving a marked point $\xi$, then
$${\partial\over\partial\xi}\langle X\rangle=
\oint_\xi {dz\over 2\pi i}\langle T(z) X\rangle.
\eqno (3.6) $$
It allows us to view any meromorphic 2-differential with poles at marked
points as a 1-form on the moduli space. For a surface
of genus $N\ge 2$ with $n$ marked points there exist
$(3N-3+n)$ linearly independent meromorphic 2-differentials with
simple poles at marked points. They form a basis in the
cotangent space to the moduli space of the surface with
marked points. Remark that the second-order residues of the
stress-energy tensor are fixed by the field dimensions,
higher orders vanish if the fields are primary with respect
to Virasoro algebra.

Due to the equivalence between 2-differentials on the surface and 1-forms
on the moduli space, the differential $d_m$ on the moduli space
mapping functions to 1-forms can be treated as an operator
mapping functions on the moduli space to 2-differentials on the surface.
We denote this operator by $d_m(z)$ so that
$$ \delta_\mu F=\int_\Sigma \mu(z,\bar z) \, d_m(z) F
\, dz d\bar z
\eqno (3.7) $$
for any function $F$ of the moduli. Notice that for $F$ being
a correlation function,
$$ d_m(z) \langle X \rangle =\langle T(z) X \rangle.
\eqno (3.8) $$

\bigskip
\centerline{4. Twisted 1-forms on the surface and}
\centerline{the twisted b-c system.}
\bigskip
\nobreak
{\it Twisted 1-forms.}

\nobreak
Any correlation function of the form $\langle j^a(z) X\rangle$ is a
meromorphic twisted 1-form in $z$ with values in the Lie algebra $\G$
(we shall also call it a twisted 1-form in $z,a$). Twisting means that
passing across A-cycles (where the twists $\hat g_i$ are placed) this
1-form is conjugated by the corresponding twists.
More formally twisted 1-forms can be described with the help of
the ``Schottky-type'' covering. Namely, let us consider the
covering $\Sigma^*$ of the surface $\Sigma$ such that the set of B-generators
of $\pi_1(\Sigma)$ lifts to an infinite tree on $\Sigma^*$, and A-generators
lift again to closed loops. Recall that the fundamental group
$\pi_1(\Sigma)$ is generated by the $2N$ elements $A_i$ and $B_i$,
$i=1,\dots,N$, with the only defining relation
$$ A_1 B_1 A_1^{-1} B_1^{-1} A_2 B_2 A_2^{-1} B_2^{-1}
\dots   A_N B_N A_N^{-1} B_N^{-1} =1.
\eqno (4.1) $$
$\pi_1(\Sigma^*)$ is the subgroup of $\Sigma$ generated by all
$A_i$ and group elements conjugate to $A_i$. Evidently, it is
a normal subgroup, and the group
$$\Gamma=\pi_1(\Sigma)/ \pi_1(\Sigma^*)
\eqno(4.2) $$
is the group of covering transformations of $\Sigma^*$, interchanging
points with the same projections to $\Sigma$. For our choice
of the covering, $\Gamma$ is the free group with $N$ generators.
Choose the generators to be the classes of $B_i$ and denote the
corresponding transformations of $\Sigma^*$ by $\gamma_i$.

This covering arises in the Schottky construction (see Appendix A).
In that case $\Gamma$ is the Schottky group, $\Sigma^*$ is
the Riemann sphere without fixed points of $\Gamma$, $\gamma_i$
are the projective transformations generating $\Gamma$.
We should emphasize though, that the construction of this covering
does not restrict our choice of parametrizations of the surface
and the moduli nor the choice of projective structures.

After introducing the covering $\Sigma^*$ we can define the twisted
1-form as a 1-form on the covering obeying proper transformation rules.
Assume that the local coordinates on different sheets of the covering
are related by the action of $\Gamma$. Then we call
$f^a(z)$ a twisted 1-form in $z,a$ if for all $i=1,\dots,N$
$$\gamma^\prime_i (z) f^a(\gamma_i(z)) = (g_i)^a_b f^b(z),
\eqno (4.3) $$
where $(g_i)^a_b$ are the matrix elements of the twists in the adjoint
representation.

{\it Main examples of meromorphic twisted 1-forms.}

\nobreak
To reconstruct twisted 1-forms from their singularities and
integrals over A-cycles, we shall decompose them in a sum
of suitable finite-dimensional spaces. For our construction
we need to fix an auxiliary point $w_0$ on the surface.

{\it (i) Twisted 1-forms $\omega^a_{ib}(z;w_0)$.}
Consider first the space of all meromorphic
twisted 1-forms with the only simple
pole at $w_0$. This space is $(N\dim \G)$-dimensional,
and any 1-form from this space is uniquely determined by
its integrals over A-cycles.
Define a basis $\omega^a_{ib}(z;w_0)$ in this space by
$$\oint_{A_i} \omega^a_{ib} (z;w_0) dz = \delta^a_b \delta_{ij}.
\eqno(4.4) $$
The residues at $z=w_0$ are
$$ \Res{w_0} \omega^a_{ib} (z;w_0)= (g_i^{-1} -1)^a_b.
\eqno(4.5) $$

In the Schottky parametrization such twisted 1-forms are given
by the Poincar\'e series
$$\omega^a_{ib}(z;w_0)=\sum_{\gamma\in\Gamma}
(g_\gamma^{-1})^a_b \left[{\gamma^\prime (z)\over \gamma(z)-
\gamma_i(w_0)} - {\gamma^\prime (z)\over \gamma(z)-w_0} \right].
\eqno (4.6) $$

{\it (ii) Twisted 1-forms  $\theta^a_b(z;w,w_0)$.}
Another useful example of twisted 1-forms are 1-forms
$\theta^a_b(z;w,w_0)$ with zero
integrals over A-cycles and a given simple pole at a given point
$w$ (which is compensated by a simple pole at the auxiliary point
$w_0$):
$$\oint_{A_i} \theta^a_b (z;w,w_0) \, dz = 0,
\eqno(4.7) $$
$$ \Res{w} \theta^a_b (z;w,w_0)=
-\Res{w_0}\theta^a_b(z;w,w_0)=\delta^a_b.
\eqno(4.8) $$

In the Schottky parametrization the Poincar\'e series
for $\theta^a_b(z;w,w_0)$ is
$$\theta^a_b(z;w,w_0)=\sum_{\gamma\in\Gamma}
(g_\gamma^{-1})^a_b \left[{\gamma^\prime (z)\over \gamma(z)-w} -
{\gamma^\prime (z)\over \gamma(z)-w_0} \right].
\eqno (4.9) $$

Here we assumed that $w$ and $w_0$ belong to the fundamental domain
of $\Gamma$. Holomorphically extending these expressions
to other sheets of the covering $\Sigma^*$, we can write
$$ \omega^a_{ib}(z;w_0)=\theta^a_b(z;\gamma_i(w),w_0).
\eqno(4.10)$$

The transformation properties of $\theta^a_b(z;w,w_0)$ as a function
of the parameters $w$ or $w_0$ are described by
$$ \theta^a_b(z;\gamma_j(w),w_0)=\theta^a_c(z;w,w_0)(g_j^{-1})^c_b
+\omega^a_{jb}(z;w_0).
\eqno (4.11) $$
The proof of this equality follows from the observation that l.h.s.
and r.h.s. are both twisted 1-forms in $z,a$ and have equal singularities
and equal integrals over A-cycles.

{\it (iii) Twisted 1-forms $\Omega^{ab}(z,w)$.}
Differentiating $\theta^a_b(z;w,w_0)$ with respect to the parameters
we obtain another twisted 1-form $\Omega^{ab}(z,w)$
with a pole of second order:
$$\Omega^{ab}(z,w)={\partial\over\partial w}\theta^a_b(z;w,w_0)=
-{\partial\over\partial w}\theta^a_b(z;w_0,w),
\eqno(4.12) $$
The first order residue of $\Omega^{ab}(z,w)$ at $z=w$ vanishes
as well as integrals over A-cycles.
A remarkable property of $\Omega^{ab}(z,w)$ is its symmetry under
interchanging $(z,a)\leftrightarrow (w,b)$:
$$ \Omega^{ab}(z,w)=\Omega^{ba}(w,z),
\eqno(4.13) $$
in particular, $\Omega^{ab}(z,w)$ is a twisted 1-form both in
$z,a$ and in $w,b$.

The Poincar\'e series for $\Omega^{ab}(z,w)$
in Schottky parametrization is
$$ \Omega^{ab}(z,w)=
\sum_{\gamma\in\Gamma}
(g_\gamma^{-1})^a_b {\gamma^\prime (z)\over (\gamma(z)-w)^2}.
\eqno(4.14) $$

The twisted 1-forms $\omega^a_{ib}(z;w_0)$ and $\theta^a_b(z;w,w_0)$
can be expressed in terms of $\Omega^{ab}(z,w)$ as
$$\omega^a_{ib}(z;w_0)=\int_{w_0}^{\gamma_i(w_0)} d\xi \,
\Omega^{ab}(z,\xi),
\eqno (4.15) $$
$$\theta^a_b(z;w,w_0)=\int_{w_0}^w d\xi \, \Omega^{ab}(z,\xi).
\eqno (4.16)$$

In our discussion we shall need to look at these forms at their
singular points. For this reason we wish to treat the twisted
1-forms we defined above as correlation functions in the twisted
b-c system of spin 1. Then the regularization of singularities will
correspond to the normal ordering of the fields.

{\it Twisted b-c system.}

\nobreak
The b-c system of spin 1 contains anticommuting
fields $b$ of spin 1 and $c$ of spin 0 (i.e. $b$ field transforms
as a 1-form and $c$ field --- as a function on the surface).
For constructing the twisted version we take
$\dim\G$ copies of b-c systems. The fields $b^a$ and $c^a$
obey the OPE
$$ b^a(z) c^b(w)={\delta^{ab}\over z-w} +o(1),
\eqno (4.17) $$
and the stress-energy tensor is given by
$$ T(z)=-:\! b^a(z) \partial c^a(z)\! : \,=
-\lim_{w \to z}\left( b^a(z) \partial c^a(w)
-{\dim\G\over (z-w)^2 } \right).
\eqno (4.18) $$

We define b-c currents by
$$ j^a(z)=f^{abc}b^b(z)c^c(z),
\eqno (4.19) $$
Then these currents form the affine algebra $\hat\G$ at the
level $2h^*$:
$$ j^a(z) j^b(w) = -2h^*{\delta^{ab} \over (z-w)^2} -{f^{abc}\over
z-w} j^c(w) + O(1).
\eqno (4.20) $$
On the handles of the surface we insert the twists
$$ \hat g_i = \exp \left(\oint_{A_i} \xi_i^a j^a(z) \, dz \right)
\eqno (4.21)$$
by the same group elements as in the WZNW model.

To eliminate zero modes of $b$ and $c$ fields we include in
correlation functions, along with the twists, the product of all $c$
fields in the auxiliary point $w_0$ and the integrals of all $b$ fields
over all A-cycles:
$$ \prod_{a=1}^{\dim\G} c^a(w_0) \prod_{i=1}^N \oint_{A_i} b^a(z) \, dz.
\eqno (4.22) $$

Similarly to the twisted WZNW model, we shall suppress in notation
both the twists (4.21) and the insertions (4.22), thus writing
$\langle X \rangle^{b-c}$ instead of $\langle X
\prod_{a=1}^{\dim\G} c^a(w_0) \prod_{i=1}^N \oint_{A_i} b^a(z) \, dz \,
\prod_{i=1}^N \hat g_i \rangle^{b-c}$.

{\it Correlation functions.}

\nobreak
Consider the normalized correlation function $\langle b^a(z)
c^b(w)\rangle^{b-c}_N$. The twists (4.21) make it a twisted 1-form
in $z,a$. Its integrals over A-cycles vanish due to the insertion of
$\prod_{i,a} \oint_{A_i} b^a(z) \, dz$. It has two poles: at
$z=w$
$$\Res{w} \langle b^a(z) c^b(w)\rangle^{b-c}_N =\delta^{ab}
\eqno (4.23) $$
and at $z=w_0$ (due to the insertion of $\prod_a c^a(w_0)$).
Therefore,
$$ \langle b^a(z) c^b(w)\rangle^{b-c}_N =\theta^a_b(z;w,w_0).
\eqno (4.24) $$
Also
$$ \Omega^{ab}(z,w)= \langle b^a(z) \partial c^b(w)\rangle^{b-c}_N
= \langle b^b(w) \partial c^a(z)\rangle^{b-c}_N.
\eqno (4.25) $$
Many-point correlation functions obey the Wick rule. For
4-point correlation functions it reads:
$$\eqalign{
 \langle b^a(z_1) b^b(z_2) c^c(z_3) c^d(z_4)\rangle^{b-c}_N =&
 \langle b^a(z_1) c^d(z_4)\rangle^{b-c}_N \,
 \langle b^b(z_2) c^c(z_3)\rangle^{b-c}_N  \cr
&\qquad - \langle b^a(z_1) c^c(z_3)\rangle^{b-c}_N \,
 \langle b^b(z_2) c^d(z_4)\rangle^{b-c}_N. }
\eqno(4.26) $$
This identity can be checked easily by comparing the transformation
properties and the singularities of the l.h.s. and the r.h.s.
We shall need this formula in further computations.

{\it Regularization.}

\nobreak
We can regularize singularities by normally ordering the fields inside
the correlation functions. Mathematically, this amounts to
discarding the negative power terms in Laurent series, i.e.
$$\Omega^{ab}(z,z)_{reg}=
\langle :\! b^a(z) \partial c^b(z)\! :\rangle^{b-c}_N =
\lim_{w\to z}\left(\Omega^{ab}(z,w)
-{\delta^{ab}\over (z-w)^2} \right),
\eqno (4.27) $$
$$\theta^a_b(z;z,w_0)_{reg}=
\langle :\! b^a(z) c^b(z)\! :\rangle^{b-c}_N =
\lim_{w\to z}\left(\theta^a_b(z;w,w_0)
-{\delta^{ab}\over z-w}\right).
\eqno(4.28) $$
This way of regularization depends on the choice of the local
coordinate. Although we shall admit such a dependence in intermediate
calculations, we can check afterwards that the
final result transforms properly under changes of coordinates.
We shall indicate the regularization by the subscript ``reg''.

{\it Stress-energy tensor and partition function.}

\nobreak
Using this notation,
$$\langle T(z) \rangle^{b-c}_N= -\Omega^{aa}(z,z)_{reg}.
\eqno (4.29) $$
If we choose a proper family of projective structures on the surface
(e.g. the projective structure of the Schottky representation),
the stress-energy tensor can be
integrated (on a covering of the moduli space)
to a holomorphic partition function $Z^{b-c}$ [6,5]:
$$ d_m(z) \log Z^{b-c} =\langle T(z)\rangle^{b-c}_N
\eqno (4.30) $$
(see also section 3). The square root $\Pi$ of this partition function
will play an important role in our discussion:
$$ Z^{b-c}=\Pi^2.
\eqno(4.31) $$
Notice that $\Pi$ depends both on the moduli of the surface
and on the twists $g_i$.

The following property of $\Pi$ follows immediately from its
definition:
$$ d_m(z)\Pi=-{1\over 2}\Pi\Omega^{aa}(z,z)_{reg}
\eqno(4.32) $$

In the Schottky representation $\Pi$ can be
computed explicitly [6]:
$$\Pi_{Schottky}=\prod_{k=1}^\infty \prod_{\gamma\in prim} \det_{adj}
(1-g_\gamma K_\gamma^k).
\eqno (4.33) $$
In (4.33) the product is taken over the primitive conjugate
classes of the Schottky group ($\gamma$ and $\gamma^{-1}$ are
elements of the same primitive class), the determinants are computed in
the adjoint representation, $K_\gamma$ is the multiplier of the
transformation $\gamma$ (see Appendix A). However, we shall not
further need any explicit expression for $\Pi$, and reproduce the
formula (4.33) only to simplify comparing with other works.

\bigskip
\centerline{5. Knizhnik-Zamolodchikov-Bernard equations.}
\bigskip
\nobreak
{\it KZB equation.}

\nobreak
Arising from the Sugawara construction (2.4), the KZB equation must
have the form
$$ \left(d_m + A\right) F=0,
\eqno (5.1) $$
where $d_m$ is the differential on the moduli space, the connection
form $A$ is an operator acting on the twisted conformal block $F$.
It seems useful to write equations for the product $(F\Pi)$ instead
of $F$, since in this case the dependence on the level $k$ reduces
to the factor $(k+h^*)^{-1}$ in front of the connection form.

Let $F=\langle\Phi_1(z_1)\dots \Phi_n(z_n)\rangle$ be a non-normalized
correlation function (conformal block), with the twists implicitly
included, $\Pi$ be the square root of the partition function of the
twisted b-c system (see section 4). Let $\theta^a_b(z;w,w_0)$,
$\omega^a_{ib}(z;w_0)$ and $\Omega^{ab}(z,w)$
be the twisted 1-forms defined in section 4,
$d_m(z)$ be the differential on the moduli space which maps
functions on the moduli to 2-differentials on the surface (see section 3).
Then the KZB equation looks like follows:
$$ \eqalign{
\Biggl( d_m(z) +{1\over 2(k+h^*)}\biggl[ \Bigl(\sum_{\alpha=1}^n
t_\alpha^a
\theta^b_a(z;\xi_\alpha,w_0)+&\sum_{i=1}^N \DL^{ia} \omega^b_{ia}
(z;w_0)\Bigr)\Bigl(\sum_{\beta=1}^n
\theta^b_c(z;\xi_\beta,w_0) t_\beta^c +\sum_{j=1}^N \omega^b_{jc}
(z;w_0)\DL^{jc}\Bigr) \cr
& +U(z)\biggr]\Biggr)(F\Pi)=0,}
\eqno (5.2)$$
where all right-invariant derivatives with respect to the twists $\DL^{ia}$
must be thought of as acting also outside the brackets (in particular,
on $F\Pi$), and the ``potential'' term
$$\eqalign{
U(z)&=h^*\Omega^{aa}(z,z)_{reg} -{1\over \Pi} \DL^{ia}
\omega^b_{ia}(z;w_0)\omega^b_{jc}(z;w_0)\DL^{jc}\Pi \cr
&= -{2h^*\over \Pi}\Bigl(d_m(z)+{1\over 2h^*}\DL^{ia}
\omega^b_{ia}(z;w_0)\omega^b_{jc}(z;w_0)\DL^{jc}\Bigr) \Pi }
\eqno (5.3) $$
is just a scalar factor.
The derivation of this equation is presented in Appendix B.
To make the structure of the equation more transparent we
need to introduce more notation. Namely, we generalize the
construction of pairing 1-forms with 0-boundaries to the
twisted setup at higher genera.

{\it Pairing 1-forms with 0-boundaries.}

\nobreak
First recall how one can pair 1-forms with 0-boundaries. Let
$\omega$ be a holomorphic 1-form on a simply connected domain.
Let $C$ be a 0-boundary on this domain with values in a linear
space $L$, i.e. a set of points $z_1,\dots,z_n$ labelled
with elements $l_1,\dots,l_n$ of $L$ such that
$$ l_1+\dots+l_n=0.
\eqno (5.4) $$
Define the pairing between $C$ and $\omega$
$$\int\limits_C \omega = \sum_{\alpha=1}^n l_\alpha
\int_{w_0}^{z_\alpha} \omega,
\eqno (5.5) $$
taking values in $L$. Here $w_0$ is an arbitrary auxiliary point
inside the domain. Condition (5.4) along with the holomorphicity of
$\omega$ makes this definition independent of the choice of $w_0$ and
pathes of integration. In other words, we integrate $\omega$ over
a 1-chain with the boundary $C$, and the result does not depend
on the particular choice of the 1-chain.

This definition obviously extends to the
case of multiply connected domains and meromorphic forms $\omega$,
if all integrals of $\omega$ along noncontractible contours and
first order residues at all poles vanish. We shall further extend
this construction to twisted 1-forms on Riemann surfaces, but first
let us look more closely at the KZ equation to motivate our
discussion.

Usually we write the KZ equation in the form (1.1) although this
expression is not quite invariant under the action of $SL(2,{\bf C})$
projective transformations. Without using invariant properties of
the conformal block $F=\langle\Phi_1(z_1)\dots \Phi_n(z_n)\rangle$
equation (1.1) gets transformed to the most general form
$$\Biggl({\partial\over\partial z_\alpha}+{1\over k + h^*}
t_\alpha^a \sum_{\beta\ne\alpha}
 \Bigl({1\over z_\alpha-z_\beta}-{1\over z_\alpha-w_0}\Bigr) t_\beta^a
 \Biggr) F =0.
\eqno (5.6)$$
Thus the equation (1.1) uses the choice of the auxiliary point $w_0=\infty$.
Of course, equations (5.6) and (1.1) are equivalent on the class of
$\G$-invariant functions
$$\Bigl(\sum_{\alpha=1}^n t_\alpha^a\Bigr) F=0,
\eqno(5.7)$$
therefore we may choose any convenient value for $w_0$. The proof of (5.7)
is obvious: we surround each point $z_\alpha$ by a contour
so that the sum of these contours is homologically equivalent to zero.
Then, integrating the current $j^a(z)$ along these contours we obtain
$$\Bigl(\sum_{\alpha=1}^n t_\alpha^a \Bigr)F=
\langle \sum_{\alpha=1}^n \oint_{z_\alpha} {dz\over 2\pi i}
j^a (z)\Phi_1(z_1)\dots \Phi_n(z_n)\rangle=0.
\eqno (5.8) $$
To make the independence of $w_0$ explicit in notation we rewrite
the equation in terms of pairing with 0-boundaries. Namely, the points
$z_\alpha$ labelled with the operators $t_\alpha^a$ form a 0-boundary with
values in the space of operators on $\G$-invariant elements
$f\in V_1\otimes\dots\otimes V_n$, where $V_\alpha$ is the
representation space of the field $\Phi_\alpha$. We denote such a
0-boundary by $C^a$. Then the KZ equations take the form
$$\Biggl[{\partial\over\partial z_\alpha}+{1\over k + h^*}
t_\alpha^a \Bigl(\int\limits_{C^a}
 {dz \over (z_\alpha-z)^2}\Bigr)_{reg}\Biggr] F =0.
\eqno (5.9)$$
Here $dz/(z_\alpha-z)^2$ is a meromorphic 1-form on the sphere with
the only pole of second order at $z=z_\alpha$ and with zero
first order residue. The subscript ``reg'' means discarding singular
powers when we integrate to the singular point $z_\alpha$:
$$ \Bigl(\int\limits_{C^a}{dz\over (z_\alpha-z)^2}\Bigr)_{reg}=
\lim_{z^\prime_\alpha \to z_\alpha}
\Bigl(\int\limits_{C^a}{dz\over (z^\prime_\alpha-z)^2}- {t^a_\alpha
\over z^\prime_\alpha-z} \Bigr).
\eqno (5.10) $$
The regularization depends on the choice of the coordinate $z$
around $z_\alpha$, which is consistent with the transformation
properties of $\Phi_\alpha$ as a conformal field with
the conformal dimension
$$\Delta_\alpha = {t_\alpha^a t_\alpha^a \over 2(k+h^*)} .
\eqno (5.11) $$

{\it Generalization to higher genera.}

\nobreak
This notation admits a generalization to higher genera. Now
the consideration analogous to (5.7) leads to the $\G$-invariance condition
$$\Bigl(\sum_{\alpha=1}^n t_\alpha^a-\sum_{i=1}^N (\DL^{ia}-\DR^{ia})
\Bigr)F(g)=0.
\eqno(5.12)$$
Here $F$ is the twisted conformal block depending on the twists $g_i$,
$N$ is the genus of the surface. $\DL^{ia}$ and $\DR^{ia}=(g_i^{-1})^a_b
\DL^{ib}$ are the right- and the left-invariant derivatives along the
twists ($(g_i^{-1})^a_b$ are the matrix
elements in the adjoint representation).
The analogue of the 0-boundary of the previous construction will be
a new object (``generalized 0-boundary'') consisting of
the points $z_a$ labelled with $t_\alpha^a$
{\it and the operators $\DL^{ia}$ labelling handles}.
Again, denote this object by $C^a$.
If $\omega^a$ is a meromorphic twisted 1-form
with vanishing first order residues and zero integrals over A-cycles,
then the operator
$$\int\limits^\to_{C^a}\omega^a = \sum_{\alpha=1}^n
\int_{w_0}^{z_\alpha} \omega^a t_\alpha^a + \sum_{i=1}^N
\int_{w_0}^{\gamma_i(w_0)} \omega^a \DL^{ia},
\eqno(5.13) $$
when restricted to $\G$-invariant functions (5.12) is independent
of the choice of $w_0$ and paths of integration (summation over $a$
is assumed).
The integrals in (5.13) are defined {\it on the covering $\Sigma^*$},
therefore there is no ambiguity in the projections of the homology
classes of integration
pathes  onto B-cycles. The arrow above the integral sign indicates that we
place differentiation operators $\DL^{ia}$ to the right, thus
they do not act on $\omega$ itself. Similarly
$$\int\limits^\gets_{C^a}\omega^a = \sum_{\alpha=1}^n
t_\alpha^a \int_{w_0}^{z_\alpha} \omega^a + \sum_{i=1}^N
\DL^{ia} \int_{w_0}^{\gamma_i(w_0)} \omega^a
\eqno(5.14) $$
is the conjugate (up to sign, since $t_\alpha^a$ and $\DL^{ia}$ are
antisymmetric) operator.

More formally, in our construction we integrate 1-forms over an
element of the operator-valued $\Gamma$-invariant relative
homology group $H^{(\Gamma)}_1(\Sigma^*,M,L)$, where $\Sigma^*$ is
the ``Schottky-type'' covering (see section 4), $M$ is the set of
preimages of the marked points on $\Sigma^*$, $L$ is the space
of operators acting on twisted conformal blocks. $\Gamma$ acts
on 1-chains by mapping in $\Sigma^*$ and simultaneously conjugating
by the corresponding twist $g_\gamma$. The element of
$H^{(\Gamma)}_1(\Sigma^*,M,L)$ as a 1-chain on $\Sigma^*$ has the
boundary consisting of the marked points with the corresponding
operators.
This condition determines the cycle in
$H^{(\Gamma)}_1(\Sigma^*,M,L)$ unambiguously after we fix the
basis of A- and B-cycles on the surface by placing twists on the
handles and specify the B-projections of the cycle to be the
differentiation operators along the twists\footnote*{The author
thanks A.Losev for elucidating this point.}.

{\it Final form of the equation.}

\nobreak
Using this notation we rewrite the KZB equations as
$$ \Biggl( d_m(z) +{1\over 2(k+h^*)}
\Bigl[\int\limits^\gets_{C^a} \Omega^{ab}(\eta,z) \, d\eta
\int\limits^\to_{C^c} \Omega^{bc}(z,\xi) \, d\xi
+U(z)\Bigr]\Biggr)(F\Pi)=0.
\eqno (5.15)$$
Now recall that $\Omega^{ab}(z,w)=\langle b^a(z)\partial c^b(w)
\rangle_N^{b-c}$
in the twisted b-c system. Therefore, we naturally define
$$ d_m(z)\Omega^{ab} (u,w)= \langle T(z) b^a(u) \partial c^b(w)
\rangle_N^{b-c}
- \langle T(z)\rangle_N \langle b^a(u) \partial c^b(w)\rangle_N^{b-c}.
\eqno (5.16) $$
Since $T=-:\! b\partial c\! :$, the Wick rule (4.26) gives
$$ d_m(z)\Omega^{ab} (u,w)= \Omega^{ac} (u,z) \Omega^{cb} (z,w).
\eqno (5.17) $$
This enables us to rewrite (5.15) as
$$ \Biggl( d_m(z) +{1\over 2(k+h^*)}
\biggl[ \biggl(d_m(z) \Bigl(\int\limits^\gets_{C^a} d\eta
\int\limits^\to_{C^b} d\xi \,\Omega^{ab} (\eta,\xi)\Bigr)_{reg}\biggr)
+U(z)\biggr]\Biggr)(F\Pi)=0.
\eqno (5.18)$$
Here the regularization is introduced to eliminate logarithmic divergences
when we perform both integrations to the same marked point $z_\alpha$:
$$\Bigl(\int\limits^\gets_{C^a} d\eta
\int\limits^\to_{C^b} d\xi \,\Omega^{ab} (\eta,\xi)\Bigr)_{reg}
= \lim_{z^\prime_\alpha \to z_\alpha}
\Bigl[\int\limits^\gets_{C^{a\prime}} d\eta
\int\limits^\to_{C^b} d\xi \,\Omega^{ab} (\eta,\xi)
- \sum_{\alpha=1}^n t_\alpha^a t_\alpha^a \log
(z^\prime_\alpha - z_\alpha)\Bigr].
\eqno (5.19) $$
Recall that a connection $\nabla=d+\lambda A$ is flat (the equations
$(\partial_\mu+\lambda A_\mu)F=0 $ are compatible) for any
$\lambda$ if and only if
$dA=0$ and $[A_\mu,A_\nu]=0$ for any directions $\mu$ and $\nu$ on
the moduli space. The first of these two conditions is almost checked
(the first summand is locally integrated), except the potential term
$U(z)$. $U(z)$ is a holomorphic 2-differential on the surface, and therefore
represents a 1-form on the moduli space (since it can be coupled to
Beltrami differentials, see section 3). Appendix C contains the proof
that this 1-form on the moduli space is closed, i.e.
$d_m(w) U(z) - d_m(z) U(w) =0$. Therefore, there locally exists
a function $W$ of the moduli and the twists such that
$$ U(z)=d_m(z) W.
\eqno (5.20) $$
We may define the ``universal'' operator
$$ A_{KZB}=\Bigl(\int\limits^\gets_{C^a} d\eta
\int\limits^\to_{C^b} d\xi \,\Omega^{ab} (\eta,\xi)\Bigr)_{reg}
+W.
\eqno (5.21) $$
Using this operator, we rewrite the KZB equations in the most invariant
form
$$ \Bigl[d_m+ {1\over 2(k+h^*)} (d_m A_{KZB})\Bigr](F\Pi)=0
\eqno (5.22) $$
for any variation of the moduli.

{\it Dependence on the coordinates.}

\nobreak
Let us discuss how the terms of this equation depend on the choice
of the coordinates on the surface. Notice that $d_m A_{KZB}$ is not quite
a 1-form on the moduli space (or, equivalently, $A_{KZB}$ is not
a function). Indeed, there are two coordinate dependent regularizations
in this term.One of them is performed in (5.19) and depends on the
local coordinates at the marked points. This dependence gives the
correct conformal dimensions (5.11) of the fields $\Phi_\alpha$.
The other regularization appears in $U(z)$ and depends on
the projective structure on the surface. This regularization
provides for $d_m A_{KZB}$ the transformation properties (3.4) of
the stress-energy tensor with the Virasoro central charge
$$ c= -{h^*\dim\G\over k+h^*} = {k\dim\G\over k+h^*} -\dim\G
=c_{{}_{WZNW}}+{1\over 2}c_{b-c},
\eqno (5.23) $$
which is exactly the central charge of $(F\Pi)$.

Remark that since $U(z)$ depends on the choice of the projective
structure, the question of whether it is zero does not make sense.
However, if $U(z)$ does not depend on the twists, then there exists
a projective structure (induced by $U(z)$) in which $U(z)=0$.
This happens in the case of torus, but we believe that at higher
genera $U(z)$ depends on the twists nontrivially.

{\it Examples.}

\nobreak
We can illustrate our construction by three simple examples. In
the case of zero genus (sphere) we have
$$\Omega^{ab}(z,w)_{sphere}={\delta^{ab}\over (z-w)^2},
\eqno (5.24)$$
and
$$ A_{KZB}^{sphere}= \sum_{\alpha\ne\beta}t_\alpha^a t_\beta^a
\log (z_\alpha-z_\beta).
\eqno (5.25)$$

In the case of torus (genus one) the potential term vanishes
in the Schottky parametrization[3]:
$$ U^{torus}_{Schottky}=0.
\eqno (5.26)$$
If there are no marked points, a torus is characterized by
a single complex moduli parameter $q$. The torus with a given $q$ is
defined as the quotient of ${\bf C}^*$ by the equivalence $z\sim qz$.
Elementary computations show that on a torus without marked points
$$A_{KZB}^{torus}=\log q \ \Delta,
\eqno (5.27) $$
where $\Delta=\DL^a \DL^a $ is the Laplacian on the group $G$ of the
twist [3].

In the abelian case ($G=U(1)$) $\Pi$ does not depend on the twists,
and the potential $U(z)$ vanishes. In this case
$$A_{KZB}^{abel}=\tau_{ij}\DL^i \DL^j,
\eqno (5.28) $$
where $\tau_{ij}$ is the period matrix of the surface:
$$ \tau_{ij}=\int_{w_0}^{\gamma_j(w_0)} \omega_i(z)\, dz
\eqno (5.29) $$
(in the abelian case $\omega^a_{ib}(z;w_0)$ becomes proportional
to $\delta^a_b$ and independent of $w_0$).

{\it Symmetricity.}

\nobreak
Our computations are consistent with Losev's observation that
the KZB connection form must be symmetric [5].
The symmetricity of $A_{KZB}$ and, therefore, of the connection form
$d_m A_{KZB}$ indicates that the connection preserves a certain pairing
between solutions of KZB equations with central charges $k$ and
$(-k-2h^*)$.

Let solutions $F_k$ and $F_{-k-2h^*}$ of KZB equations
for central charges $k$ and $(-k-2h^*)$ respectively take values in
tensor products $V_1\otimes\dots\otimes V_n$
and $V_1^*\otimes\dots\otimes V_n^*$ of representations
of $G$ (here $V_\alpha^*$ is the representation dual to $V_\alpha$).
Then we can define a pairing between
$F_k$ and $F_{-k-2h^*}$ by
$$\eqalign{
(F_k,F_{-k-2h^*})^{b-c} &=
\int dg \,\Pi^2(g) \langle F_k,F_{-k-2h^*} \rangle \cr
&=\int \prod_{i=1}^N dg_i \,\Pi^2(g) F_k^{i_1 \dots i_N} (g)
F_{-k-2h^*}^{j_1 \dots j_N} (g) \eta^{(1)}_{i_1 j_1} \dots
\eta^{(N)}_{i_N j_N}, }
\eqno (5.30) $$
where $\eta^{(\alpha)}_{ij}$ are the matrices of the pairings
$V_\alpha\otimes V_\alpha^* \to {\bf C}$, the integration is
performed over all twists,
$dg_i$ is the invariant measure on $G$.
With respect to this bilinear form operators $t_\alpha^a$, $\DL^{ia}$
and $\DR^{ia}$ are antisymmetric, therefore $A_{KZB}$ and
$d_m A_{KZB}$ are symmetric. Hence, for the solutions $F_k$ and
$F_{-k-2h^*}$ of the KZB equations we have
$$\eqalign{
& d_m (F_k, F_{-k-2h^*})^{b-c} =
(d_m F_k, F_{-k-2h^*})^{b-c} +
(F_k, d_m F_{-k-2h^*})^{b-c}  \cr
& = {1\over k+h^*} \Bigl( (d_m A_{KZB}) F_k, F_{-k-2h^*}\Bigr)^{b-c} +
{1\over (-k-2h^*)+h^*} \Bigl( F_k,
(d_m A_{KZB}) F_{-k-2h^*}\Bigr)^{b-c} = 0 }
\eqno (5.31) $$
due to the symmetricity of $d_m A_{KZB}$.
Losev [5] interpreted the symmetricity of the connection
as a consequence of an operator
formalism in $G/G$ WZNW coset theory. Referring to the operator formalism
we could treat $F_{-k-2h^*}$ as a conformal block with the central
charge $(-k-2h^*)$, and the twisted b-c system --- as ghosts in
BRST construction [5].

{\it Integrability.}

\nobreak
Now let us return to the integrability conditions. To explicitly
prove the flatness of the connection it remains to show that
$$ [\partial_\mu A_{KZB},\partial_\nu A_{KZB}]=0
\eqno (5.32) $$
for any directions $\mu$ and $\nu$ in the moduli space.
In Appendix D we prove this for $\partial_\mu$ and $\partial_\nu$ being
derivatives with respect to the coordinates of inserted fields. We also
derive the corresponding generalization of the classical
Yang-Baxter equation,
which though does not seem to be very instructive (see Appendix D).
Instead of commuting the most general operators with marked points,
in Appendix E we prove the compatibility  of the equations for the
partition function (with no marked points). We claim that since
marked points can be obtained in a degeneration of the surface
(double points), we indeed proved the general form of the statement.
The compatibility of the equations for the setup with marked points
(and the result of  Appendix D in particular) can be obtained as a
limiting case of the statement of Appendix E (we present an independent
derivation in Appendix D just to study the generalization of the
classical Yang-Baxter equation). Thus we proved that
the KZB connection is flat. Remark that our proof did not refer to
any particular projective structure; the only property we used was
that the holomorphic b-c partition function exists (i.e. the
projective structure is ``compatible'' --- see section 3).

\bigskip
\centerline{6. Conclusion.}
\bigskip
\nobreak
To summarize, we have written the KZB equations and proved their
integrability. The concise form (5.21), (5.22) of the equations
contrasts sharply with our way of proving the compatibility
(Appendices C,D,E). We still hope that the structure of the equations and
their relation to the b-c systems can suggest a more elegant treatment
of the problem and help to avoid the tedious computations. Another
question remains open about the meaning of the potential term.
One may look for its geometric description or for its expression
in the WZNW coset construction [5]. We also admit the possibility
to explicitly integrate the potential $U(z)$ to $W$ (see equation
(5.20)). Another approach to the problem can be inspired
by the works on geometric
quantization of the moduli space of flat connections, which arrive
to similar equations [8].

\bigskip
\centerline{Acknowledgements.}
\bigskip
\nobreak
The author wishes to thank A.Losev for initiating this
research and friendly guidance throughout the work,
A.Morozov, N.Nekrasov and V.Fock for stimulating discussions.
The author acknowledges the support of the Russian
Foundation of Fundamental Research under the Grant No.93-02-14365.

\bigskip
\centerline{Appendix A. Schottky parametrization of Riemann surfaces.}
\bigskip
\nobreak
In the Schottky parametrization the surface is constructed as the
quotient of the Riemann sphere (more strictly, of the sphere without
the fixed points of the group) by the action of a Schottky group.
The Schottky group $\Gamma$ is a group freely generated by $N$
projective maps $\gamma_i$ such that one can find $2N$
circles $A_i$ and $A_i^\prime=\gamma_i(A_i)$, $i=1,\dots,N$ --- all
external to each other, and $\gamma_i$ maps the exterior of $A_i$
onto the interior of $A_i^\prime$.
The exterior to all the circles $A_i$ and
$A_i^\prime$ is a fundamental domain of $\Gamma$. The surface is
obtained by gluing each circle $A_i^\prime $ to $A_i$ by the
action of $\gamma_i$. Then $A_i$ become A-cycles on the surface.

For future use we should introduce two more definitions. The first
one is the parametrization of a projective transformation
$\gamma$ by its fixed points $u_\gamma$, $v_\gamma$ (repulsive
and attractive) and its multiplier $K_\gamma$ defined by
$${\gamma(z)-u_\gamma \over \gamma(z)-v_\gamma}
=K_\gamma{z-u_\gamma \over z-v_\gamma}, \qquad |K_\gamma|<1.
\eqno (A.1) $$

Finally, we shall need to extend the twists introduced on the handles
of the surface to the group homomorphism between $\Gamma$
and $G$:
$$\eqalign {
& g : \Gamma \to G, \cr
& g_{\gamma_i} = g_i,  \cr
& g_{\gamma\mu} = g_\gamma g_\mu.   }
\eqno (A.2) $$

The global coordinate $z$ on the Riemann sphere defines naturally a
projective structure on the surface. It is remarkable that in this
family of projective structures the stress-energy tensor of b-c systems
and of the WZNW model can be integrated to partition functions
(or conformal blocks) according to (3.3); i.e. in Schottky
parametrization the KZB connection becomes flat.

\bigskip
\centerline{Appendix B. Derivation of the KZB equations.}
\bigskip
\nobreak
To derive the KZB equations we first express correlation functions
with inserted currents in terms of the twisted 1-forms
$\theta^a_b(z;w,w_0)$, $\omega^a_{ib}(z;w_0)$, $\Omega^{ab}(z,w)$,
introduced in section 4. Let $\Phi$ denote the product
$\Phi_1(\xi_1)\dots\Phi_n(\xi_n)$. Then
$$\langle j^a(z)\Phi\rangle=\Bigl(\sum_\alpha \theta^a_b(z;\xi_\alpha,w_0)
t^b_\alpha + \sum_i \omega^a_{ib}(z;w_0) \DL^{ib} \Bigr)
\langle\Phi\rangle,
\eqno (B.1) $$
since r.h.s. and l.h.s. are both twisted 1-forms in $z,a$, have
equal singularities at the poles $z=\xi_\alpha$ and equal integrals
over A-cycles.
Although each of the two terms on the r.h.s. depends on the
choice of the auxiliary point $w_0$ and has a pole at $z=w_0$, their
sum does not. For further convenience we introduce more notation:
$$L^a(z;w_0)=\omega^a_{ib}(z;w_0)\DL^{ib},
\eqno (B.2) $$
$$\bar L^a(z;w_0) =\DL^{ib}\omega^a_{ib}(z;w_0)=
L^a(z;w_0)+\{\DL^{ib}\omega^a_{ib}(z;w_0)\}.
\eqno (B.3) $$
Here we use braces to specify the range of action for
the differentiation operators $\DL^{ia}$. By convention, all
differentiation operators act only inside the
braces. Also the summation over all repeating indices
(including those labelling marked points and handles) is assumed
if not specified otherwise.

For the correlation function with the insertion of two currents
we have
$$\eqalign{
\langle j^a(z) j^b(w) \Phi \rangle =& - k\Omega^{ab}(z,w)
\langle \Phi \rangle - \theta^a_d(z;w,w_0) f^{dbc}
\langle j^c(w) \Phi\rangle    \cr
& + \theta^a_c(z;\xi_\alpha,w_0)
t^c_\alpha \langle j^b(w)\Phi\rangle + L^a(z;w_0)
\langle j^b(w)\Phi\rangle      }
\eqno (B.4) $$
for the same reason of the equality of singularities and integrals
over A-cycles. Using (B.1) this becomes
$$\eqalign{
\langle j^a(z) j^b(w) \Phi \rangle =& - k\Omega^{ab}(z,w)
\langle \Phi \rangle   \cr
&- \theta^a_d(z;w,w_0) f^{dbc} \theta^c_e(w;\xi_\alpha,w_0)
t^e_\alpha \langle\Phi\rangle   \cr
&- \theta^a_d(z;w,w_0) f^{dbc}
L^c(w;w_0) \langle\Phi\rangle   \cr
&+ \theta^a_c(z;\xi_\alpha,w_0) \theta^b_d(w;\xi_\beta,w_0)
t^c_\alpha t^d_\beta \langle\Phi\rangle  \cr
&+ \theta^a_c(z;\xi_\alpha,w_0)
t^c_\alpha L^b(w;w_0)\langle\Phi\rangle  \cr
&+ L^a(z;w_0) \theta^b_c(w;\xi_\beta,w_0) t^c_\beta
\langle\Phi\rangle   \cr
&+ L^a(z;w_0) L^b(w;w_0)
\langle\Phi\rangle.    }
\eqno (B.5) $$
Introduce
$$ \Xi^a(z;w_0)=f^{abc} \theta^b_c(z;z,w_0)=
\langle j^a(z) \rangle^{b-c}_N.
\eqno (B.6) $$
Then, taking the limit $w\to z$,
$$\eqalign{
\langle :\! j^a(z) j^a(z)\! : \Phi \rangle =& - k
\Omega^{aa}(z,z)_{reg} \langle \Phi \rangle   \cr
&+  \Xi^a(z;w_0)\theta^a_e(z;\xi_\alpha,w_0)
t^e_\alpha \langle\Phi\rangle   \cr
&+ \Xi^a(z;w_0) L^a(w;w_0) \langle\Phi\rangle   \cr
&+ \theta^a_c(z;\xi_\alpha,w_0) \theta^a_d(z;\xi_\beta,w_0)
t^c_\alpha t^d_\beta \langle\Phi\rangle  \cr
&+ \theta^a_c(z;\xi_\alpha,w_0)
t^c_\alpha L^a(z;w_0)\langle\Phi\rangle  \cr
&+ L^a(z;w_0) \theta^a_c(z;\xi_\beta,w_0) t^c_\beta
\langle\Phi\rangle   \cr
&+ L^a(z;w_0) L^a(z;w_0)
\langle\Phi\rangle.    }
\eqno (B.7) $$

Now we use Losev's observation [5] that
$$\Xi^a(z;w_0) =\{2L^a(z;w_0) \log\Pi +
\DL^{ib} \omega^a_{ib}(z;w_0)\}.
\eqno (B.8) $$
The proof is based on checking that r.h.s. and l.h.s. are
twisted 1-forms with additive B-periods with equal jumps at
A-cycles and integrals over A-cycles.

This allows us to rewrite (B.7) as
$$\eqalign{
&\langle :\! j^a(z) j^a(z)\! : \Phi \rangle = \Bigl[ -k
\Omega^{aa}(z,z)_{reg}  \cr
&+ \Bigl(\bar L^a(z;w_0)+t_\alpha^b \theta^a_b(z;\xi_\alpha,w_0)\Bigr)
\Bigl(L^a(z;w_0)+t_\beta^c \theta^a_c(z;\xi_\beta,w_0)\Bigr)  \cr
&+ 2\Bigl\{L^a(z;w_0)\log\Pi\Bigr\}
\Bigl(L^a(z;w_0)+t_\beta^c \theta^a_c(z;\xi_\beta,w_0)\Bigr)\Bigr]
\langle\Phi\rangle.    }
\eqno (B.9) $$
By the Sugawara construction (2.4) and the identity (4.32) we have
$$\eqalign{
 d_m(z)\Bigl(\langle\Phi\rangle\Pi\Bigr)=&-{1\over 2(k+h^*)}
\Bigl[h^*\Pi\Omega^{aa}(z,z)_{reg}   \cr
&+ \Pi \Bigl(\bar L^a(z;w_0)+t_\alpha^b \theta^a_b(z;\xi_\alpha,w_0)\Bigr)
\Bigl(L^a(z;w_0)+t_\beta^c \theta^a_c(z;\xi_\beta,w_0)\Bigr)  \cr
&+ 2\{L^a(z;w_0)\Pi\}
\Bigl(L^a(z;w_0)+t_\beta^c \theta^a_c(z;\xi_\beta,w_0)\Bigr)\Bigr]
\langle\Phi\rangle    \cr
=&-{1\over 2(k+h^*)} \Biggl(\Bigl[h^*\Pi\Omega^{aa}(z,z)_{reg}   \cr
&+ \Bigl(\bar L^a(z;w_0)+t_\alpha^b \theta^a_b(z;\xi_\alpha,w_0)\Bigr)
\Bigl(L^a(z;w_0)+t_\beta^c \theta^a_c(z;\xi_\beta,w_0)\Bigr)\Bigr]
\Bigl(\langle\Phi\rangle\Pi\Bigr)   \cr
&- \langle\Phi\rangle \bar L^a(z;w_0) L^a(z;w_0) \Pi\Biggr)  \cr
=-{1\over 2(k+h^*)}
\Biggl(\Bigl(&\bar L^a(z;w_0)+t_\alpha^b \theta^a_b(z;\xi_\alpha,w_0)\Bigr)
\Bigl(L^a(z;w_0)+t_\beta^c \theta^a_c(z;\xi_\beta,w_0)\Bigr) + U(z)\Biggr)
\Bigl(\langle\Phi\rangle\Pi\Bigr),    }
\eqno (B.10) $$
where
$$ U(z)=\{h^*\Omega^{aa}(z,z)_{reg} -
{1\over \Pi} \bar L^a(z;w_0) L^a(z;w_0) \Pi \}.
\eqno (B.11) $$

\bigskip
\centerline{Appendix C. ``Potential'' term is a closed 1-form}
\centerline{on the moduli space.}
\bigskip
\nobreak
Closeness of the potential term $U(z)$
is equivalent to
$$d_m(w) U(z) - d_m(z) U(w) = 0,
\eqno (C.1) $$
where the operators $d_m(z)$ should be understood as in (5.16), (5.17).
Before proving this let us derive several useful identities. We
shall use the notation introduced in Appendix B.

First, we wish to prove that
$$\eqalign{
& L^a(z;w_0)\Omega^{bc}(w,\xi)-L^b(w;w_0)\Omega^{ac}(z,\xi) \cr
& +\Bigl[\Omega^{ad}(z,\xi)\theta^b_e(w;\xi,w_0)-
\Omega^{bd}(w,\xi)\theta^a_e(z;\xi,w_0)\Bigr] f^{cde} \cr
& +\Omega^{ec}(w,\xi)\theta^a_d(z;w,w_0) f^{bde}
  -\Omega^{ec}(z,\xi)\theta^b_d(w;z,w_0) f^{ade} =0 }
\eqno (C.2) $$
The proof consists of the following steps:
 (i) Observe that l.h.s. is a twisted 1-form in $z,a$. To check this
we use (4.11) along with the identity
$$ g t^a g^{-1} = (g)^b_a t^b
\eqno (C.3) $$
in the adjoint representation.
 (ii) L.h.s. is regular at $z=w$.
 (iii) L.h.s. is regular at $z=\xi$.
 (iv) Integrals over A-cycles vanish. Assuming that $w$, $\xi$, $w_0$
belong to the fundamental domain,
$$\oint_{A_i} (l.h.s.) dz = \DL^{ia} \Omega^{bc}(w,\xi)
- \oint_{A_i} \Omega^{ec}(z,\xi) \theta^b_d(w;z,w_0) f^{ade} dz.
\eqno (C.4) $$
We can prove that this is identically zero by considering correlation
functions in the twisted b-c system:
$$ \eqalign{
& \DL^{ia} \Omega^{bc}(w,\xi) = \DL^{ia} \langle b^b(w)
\partial c^c(\xi)\rangle^{b-c}_N  \cr
&=\langle\oint_{A_i} j^a(z)\, dz\ b^b(w)\partial c^c(\xi)
\rangle^{b-c}_N -\langle\oint_{A_i} j^a(z)\, dz\rangle^{b-c}_N \,
\langle b^b(w)\partial c^c(\xi)
\rangle^{b-c}_N. }
\eqno (C.5) $$
Recalling that $j^a(z)=f^{abc} b^b(z) c^c(z)$ and using the Wick rule,
we arrive to
$$ \eqalign{
\DL^{ia} \Omega^{bc}(w,\xi) &=
- f^{ade} \oint_{A_i} dz \langle b^b(w) c^e(z) \rangle^{b-c}_N
\langle b^d(z)\partial c^c(\xi) \rangle^{b-c}_N  \cr
&= - f^{ade} \oint_{A_i} dz \Omega^{dc}(z,\xi) \theta^b_e(w;z,w_0), }
\eqno (C.6) $$
which finishes the proof.

We can obtain more identities by integrating (C.2) in $\xi$ along
different pathes. If $\xi$ runs from $w_0$ to $u$, we have
$$\eqalign{
& L^a(z;w_0)\theta^b_c(w;u,w_0)-L^b(w;w_0)\theta^a_c(z;u,w_0) \cr
& +\theta^a_d(z;u,w_0)\theta^b_e(w;u,w_0) f^{cde} \cr
& +\theta^e_c(w;u,w_0)\theta^a_d(z;w,w_0) f^{bde}
  -\theta^e_c(z;u,w_0)\theta^b_d(w;z,w_0) f^{ade} =0 }
\eqno (C.7) $$
Setting $u=\gamma_i(w_0)$, this easily leads to the commutation
relation
$$ [L^a(z;w_0), L^b(w;w_0)] =
f^{acd}\theta^b_c(w;z,w_0)L^d(z;w_0) -
f^{bcd}\theta^a_c(z;w,w_0)L^d(w;w_0).
\eqno (C.8) $$

Let us now return back to proving (C.1). Since we are interested
in the antisymmetric with respect to interchanging
$z\leftrightarrow w$ part, we shall use the symbol $\simeq$ to
indicate that the antisymmetric parts of the two expressions are
equal.
$$ d_m(z) U(w)=h^* d_m(z) \Omega^{aa}(w,w)_{reg} -
{1\over 2} d_m(z) \Bigl({1\over\Pi}\bar L^a(w;w_0) L^a(w;w_0) \Pi \Bigr).
\eqno (C.9) $$
The first summand has zero antisymmetric part, since
$\Omega^{aa}(w,w)_{reg}=-2 d_m(w) \log\Pi $ is an exact 1-form
on the moduli space, and $d_m^2=0$. Also, since
$$ \bar L^a(w;w_0) L^a(w;w_0) = d_m(w)
\Bigl(\DL^{ib}\int_{w_0}^{\gamma_i(w_0)} d\xi
\int_{w_0}^{\gamma_j(w_0)} d\eta \,\Omega^{bc}(\xi,\eta) \DL^{jc} \Bigr),
\eqno (C.10) $$
we arrive to
$$ \eqalign{
d_m(z) U(w) & \simeq -{1\over 2}\Bigl[\Bigl(d_m(z){1\over\Pi}\Bigr)
\bar L^b(w;w_0) L^b(w;w_0)\Pi +
{1\over\Pi}\bar L^b(w;w_0) L^b(w;w_0)\Bigl(d_m(z)\Pi\Bigr)\Bigr] \cr
&= {1\over 4} \Bigl[- {\Omega^{aa}(z,z)_{reg} \over\Pi}
\bar L^b(w;w_0) L^b(w;w_0) \Pi + {1\over\Pi}
\bar L^b(w;w_0) L^b(w;w_0) \Bigl(\Pi\Omega^{aa}(z,z)_{reg}\Bigr)\Bigr] \cr
&= {1\over 4} \Bigl[{2\over\Pi} \Bigl\{L^b(w;w_0)
\Omega^{aa}(z,z)_{reg} \Bigr\}
\Bigl\{L^b(w;w_0) \Pi \Bigr\} + \bar L^b(w;w_0) L^b(w;w_0)
\Omega^{aa}(z,z)_{reg}\Bigr] \cr
&= {1\over 4} \Bigl[L^b(w;w_0) L^b(w;w_0) \Omega^{aa}(z,z)_{reg}
+\Xi^b(w;w_0) L^b(w;w_0) \Omega^{aa}(z,z)_{reg}\Bigr]    }
\eqno (C.11) $$
{}From (C.2) we deduce that
$$ L^b(w;w_0) \Omega^{aa}(z,z)_{reg} =
L^a(z;w_0) \Omega^{ba}(w,z) +
\Bigl[\Omega^{ea}(w,z) \theta^a_d(z;w,w_0) f^{bde} +
\Omega^{bd}(w,z) \Xi^d(z;w_0) \Bigr].
\eqno (C.12) $$
Using it in rewriting the first summand in (C.11), we obtain
$$\eqalign{
& 4 d_m(z) U(w) \simeq L^b(w;w_0) L^a(z;w_0) \Omega^{ab}(z,w) \cr
& + L^b(w;w_0) \Bigl[\Omega^{ea}(w,z) \theta^a_d(z;w,w_0) f^{bde}
+\Omega^{bd}(w,z) \Xi^d(z;w_0) \Bigr] -
\Xi^b(z;w_0) L^b(z;w_0)\Omega^{aa}(w,w)_{reg} }
\eqno (C.13) $$
Commuting $L^a(z;w_0)$ and $L^b(w;w_0)$ in the first term according to
(C.8),
$$ \eqalign{
 4 d_m(z) U(w) & \simeq f^{bcd}
 \theta^a_c(z;w,w_0) L^d(w;w_0)\Omega^{ab}(z,w) -
 f^{edb} \theta^a_d(z;w,w_0) L^b(w;w_0)\Omega^{ae}(z,w) \cr
 &+ \Omega^{ea}(w,z) L^b(w;w_0) \theta^a_d(z;w,w_0) f^{bde} \cr
 &+ \Omega^{bd}(w,z) L^b(w;w_0) \Xi^d(z;w_0) \cr
 &- \Xi^b(z;w_0) \Bigl[ L^b(z;w_0) \Omega^{aa}(w,w)_{reg}
 - L^a(w;w_0)\Omega^{ba}(z,w)\Bigr]  \cr
 &= \Omega^{ea}(w,z) L^b(w;w_0) \theta^a_d(z;w,w_0) f^{bde} \cr
 &+ \Omega^{bd}(w,z) L^b(w;w_0) \Xi^d(z;w_0) \cr
 &- \Xi^b(z;w_0) \Omega^{ea}(z,w) \theta^a_d(w;z,w_0) f^{bde}  \cr
 &- \Xi^b(z;w_0) \Omega^{bd}(z,w) \Xi^d(w;w_0).  }
\eqno (C.14) $$
The last term is symmetric, therefore,
$$ \eqalign{
 4 d_m(z) U(w) & \simeq
 \Omega^{ea}(w,z) L^b(w;w_0) \theta^a_d(z;w,w_0) f^{bde} \cr
 &- \Omega^{ea}(w,z) L^a(z;w_0) \theta^b_d(w;w,w_0) f^{bde}\cr
 &- \Xi^b(z;w_0) \Omega^{ea}(z,w) \theta^a_d(w;z,w_0) f^{bde}  \cr
 &=\Omega^{ea}(w,z)
 \Bigl[\theta^a_f(z;w,w_0) \theta^b_g(w;w,w_0)_{reg} f^{dfg} \cr
 &+ \theta^g_d(w;w,w_0)_{reg} \theta^a_f(z;w,w_0) f^{bfg}
 -  \theta^g_d(z;w,w_0) \theta^b_f(w;z,w_0) f^{afg} \Bigr]   \cr
 &- \Xi^b(z;w_0) \Omega^{ea}(z,w) \theta^a_d(w;z,w_0) f^{bde}  \cr
 &\simeq \Omega^{ea}(w,z) \theta^a_f(z;w,w_0)
 \theta^b_g(w;w,w_0)_{reg} f^{bde} f^{dfg}  \cr
 &+ \Omega^{ea}(w,z) \theta^g_d(w;w,w_0)_{reg}
 \theta^a_f(z;w,w_0) \Bigl[-f^{bfd} f^{beg} - f^{bef} f^{bdg}\Bigr] \cr
 &+ \Omega^{ea}(w,z) \Xi^b(w;w_0) \theta^a_d(z;w,w_0) f^{bde} \cr
 &=0.  }
\eqno (C.15) $$
We used (C.7) and the Jacobi identity
$$ f^{abc} f^{ade} + f^{abd} f^{aec} + f^{abe} f^{acd} = 0.
\eqno (C.16) $$

\bigskip
\centerline{Appendix D. Compatibility of the KZB equations}
\centerline{for moving marked points.}
\bigskip
\nobreak
We check the compatibility of the equations for the derivatives of the
conformal block with respect to the coordinates of marked points.
It is slightly simplifies computations if we deal with the equations
for ``bare'' conformal blocks
$F=\langle\Phi_1(\xi_1)\dots\Phi_n(\xi_n)\rangle$, not for
$(\Pi F)$. The equations for $F$ look like follows:
$$ \Biggl[{\partial\over\partial\xi_\alpha} + {1\over k+h^*} t_\alpha^a
\Bigl(\theta^a_b(\xi_\alpha;\xi_\beta,w_0)_{(reg)} t_\beta^b
+L^a(\xi_\alpha;w_0)\Bigr)\Biggr] F = 0.
\eqno (D.1) $$
The regularization of $\theta^a_b(\xi_\alpha;\xi_\beta,w_0)$ is assumed
when $\beta=\alpha$.
Our goal is to show that the $n$-point operators
$$ A_\alpha^{(n)} =  t_\alpha^a
\Bigl(\theta^a_b(\xi_\alpha;\xi_\beta,w_0)_{(reg)} t_\beta^b
+L^a(\xi_\alpha;w_0)\Bigr)
\eqno (D.2) $$
commute:
$$ [A_\alpha^{(n)}, A_\beta^{(n)}]=0.
\eqno (D.3) $$
The commutativity property is sufficient
to check only for 2- and 3-point operators. The structure
of the operators $A_\alpha^{(n)}$ ensures that any $n$-point
operators will then also commute. One easily observes this property
for the KZ equations (1.1). In this case the commutativity of
the 2-point operators is trivial, since
$$A_\alpha^{(2)sphere}=-A_\beta^{(2)sphere}
\eqno (D.4) $$
for two marked points $\alpha$ and $\beta$. The 3-point operators
commute due to the classical Yang-Baxter equation
$$ [R_{\alpha\beta}^{sphere}, R_{\beta\gamma}^{sphere}]
+[R_{\beta\gamma}^{sphere}, R_{\gamma\alpha}^{sphere}]
+[R_{\gamma\alpha}^{sphere}, R_{\alpha\beta}^{sphere}]=0,
\eqno (D.5) $$
where the R-matrix is
$$R_{\alpha\beta}^{sphere}={t^a_\alpha t^a_\beta \over
\xi_\alpha - \xi_\beta}.
\eqno (D.6) $$

This argument slightly changes in the case of higher genera. Let
$$S_\alpha= t_\alpha^a L^a(\xi_\alpha;w_0) +
\theta^a_b(\xi_\alpha;\xi_\alpha,w_0)_{reg} t_\alpha^a t_\alpha^b,
\eqno (D.7) $$
$$R_{\alpha\beta}=
\theta^a_b(\xi_\alpha;\xi_\beta,w_0) t_\alpha^a t_\beta^b
\eqno (D.8) $$
(no summation over $\alpha$ and $\beta$). Then
$$ A_\alpha^{(n)}=S_\alpha + \sum_{\beta\ne\alpha}
R_{\alpha\beta}.
\eqno (D.9) $$
The commutativity of the 2-point operators implies
$$ [S_\alpha+R_{\alpha\beta}, S_\beta+R_{\beta\alpha}]=0.
\eqno (D.10) $$
For the 3-point operators the commutativity means
$$ [A_\alpha^{(2)}+R_{\alpha\gamma},
A_\beta^{(2)}+R_{\beta\gamma}]=0,
\eqno (D.11) $$
where $A_\alpha^{(2)}$ and $A_\beta^{(2)}$ are the 2-point
operators for the marked points $\xi_\alpha$ and $\xi_\beta$.
Up to (D.10) this is equivalent to
$$ [A_\alpha^{(2)},R_{\beta\gamma}]+[R_{\alpha\gamma},A_\beta^{(2)}]
+ [R_{\alpha\gamma}, R_{\beta\gamma}] =0
\eqno (D.12) $$
or being rewritten in another way:
$$ [R_{\alpha\beta}, R_{\beta\gamma}]
+[R_{\alpha\gamma}, R_{\beta\gamma}]
+[R_{\alpha\gamma}, R_{\beta\alpha}]=
[R_{\beta\gamma},S_\alpha] + [S_\beta,R_{\alpha\gamma}].
\eqno (D.13) $$
This is a generalization of the classical Yang-Baxter
equation (D.5) to higher genera. Notice that here we do not have
the symmetry $R_{\alpha\beta}=-R_{\beta\alpha}$ any more.

Our proof of the compatibility condition (D.3) will proceed in two
steps. First we check (D.13), then (D.10). The generalized classical
Yang-Baxter equation (D.13) follows from (C.7):
$$\eqalign{
& [R_{\alpha\beta}, R_{\beta\gamma}]
+[R_{\alpha\gamma}, R_{\beta\gamma}]
+[R_{\alpha\gamma}, R_{\beta\alpha}]
-[R_{\beta\gamma},S_\alpha] - [S_\beta,R_{\alpha\gamma}] \cr
&=t_\alpha^a t_\beta^b t_\gamma^c
\Bigl\{\theta^a_d(\xi_\alpha;\xi_\gamma,w_0)
\theta^b_e(\xi_\beta;\xi_\gamma,w_0)
f^{cde} \cr
& +\theta^a_d(\xi_\alpha;\xi_\beta,w_0)\theta^e_c(\xi_\beta;\xi_\gamma,w_0)
f^{bde}
  -\theta^b_d(\xi_\beta;\xi_\alpha,w_0)\theta^e_c(\xi_\alpha;\xi_\gamma,w_0)
f^{ade} \cr
& +L^a(\xi_\alpha;w_0)\theta^b_c(\xi_\beta;\xi_\gamma,w_0)
-L^b(\xi_\beta;w_0)\theta^a_c(\xi_\alpha;\xi_\gamma,w_0) \Bigr\}  \cr
& =0.  }
\eqno (D.14) $$
Commuting the 2-point operators, obtain
$$ [S_\alpha+R_{\alpha\beta}, S_\beta+R_{\beta\alpha}]=
K_{ic}^{ab} t_\alpha^a t_\beta^b \DL^{ic} +
X^{abc} t_\alpha^a t_\alpha^b t_\beta^c
+ \bar X^{abc} t_\beta^a t_\beta^b t_\alpha^c,
\eqno (D.15) $$
where
$$ \eqalign{
K_{ic}^{ab} &=f^{ade}
\theta^b_e(\xi_\beta;\xi_\alpha,w_0)\omega^d_{ic}(\xi_\alpha;w_0)
-f^{bde}
\theta^a_e(\xi_\alpha;\xi_\beta,w_0)\omega^d_{ic}(\xi_\beta;w_0) \cr
&+ \omega^a_{id}(\xi_\alpha;w_0)\omega^b_{ie}(\xi_\beta;w_0) f^{cde} \cr
&+ L^a(\xi_\alpha;w_0)\omega^b_{ic}(\xi_\beta;w_0)
- L^b(\xi_\beta;w_0)\omega^a_{ic}(\xi_\alpha;w_0) \cr
&=0, }
\eqno (D.16) $$
$$ \eqalign{
&X^{abc}(\xi_\alpha,\xi_\beta)= -\bar X^{abc}(\xi_\beta,\xi_\alpha)  \cr
&=f^{bde} \theta^a_d(\xi_\alpha;\xi_\alpha,w_0)_{reg}
\theta^c_e(\xi_\beta;\xi_\alpha,w_0)
+ f^{ade}
\theta^d_b(\xi_\alpha;\xi_\alpha,w_0)_{reg}
\theta^c_e(\xi_\beta;\xi_\alpha,w_0) \cr
& -L^c(\xi_\beta ;w_0)\theta^a_b(\xi_\alpha;\xi_\alpha,w_0)_{reg}
+  L^a(\xi_\alpha;w_0)\theta^c_b(\xi_\beta ;\xi_\alpha,w_0)  \cr
&+ f^{cde} \theta^a_d(\xi_\alpha;\xi_\beta,w_0)
\theta^e_b(\xi_\beta;\xi_\alpha,w_0)  \cr
& =0  }
\eqno (D.17) $$
again due to (C.7). This completes our proof of (D.3).

\bigskip
\centerline{Appendix E. Compatibility of the equations}
\centerline{for the partition function.}
\bigskip
\nobreak
We explicitly show that in the case of no insertions (equations for
the partition function) the KZB equations are compatible (the
connection is flat). This follows from
$$ [ d_m(z) A_{KZB},  d_m(w) A_{KZB} ] = 0.
\eqno (E.1) $$
We prove it in this Appendix.

Let
$$ A(z)= d_m(z) A_{KZB} = \Delta_B (z) + U(z),
\eqno (E.2) $$
where
$$ \Delta_B (z)=\bar L^a(z;w_0) L^a(z;w_0)
\eqno (E.3) $$
is a symmetric second order differential operator, $U(z)$ is
the potential term,
$$ \eqalign{
U(z) &=h^*\Omega^{aa}(z,z)_{reg}-\{ {1\over\Pi}\Delta_B(z)\Pi \}  \cr
&=h^*\Omega^{aa}(z,z)_{reg}-\{ \Delta_B(z)\log\Pi \}
-\{L^a(z) \log\Pi\}^2.  }
\eqno (E.4) $$
{}From now on we shall omit the dependence on the auxiliary point $w_0$ in
all expressions, so that $L^a(z)\equiv L^a(z;w_0)$, etc.
Commuting $A(z)$ and $A(w)$, we obtain
$$\eqalign{
[A(z), A(w)] =&[\Delta_B(z), \Delta_B(w)] + [\Delta_B(z), U(w)]
-[\Delta_B(w), U(z)]  \cr
=& [\Delta_B(z), \Delta_B(w)] + \Bigl\{L^a(z) U(w)\Bigr\} L^a(z)
+\bar L^a(z) \Bigl\{L^a(z) U(w)\Bigr\}     \cr
&- \Bigl\{L^a(w) U(z)\Bigr\} L^a(w) - \bar L^a(w) \Bigl\{L^a(w) U(z)\Bigr\}.
}
\eqno (E.5) $$
We wish to prove this to be zero.

First compute
$$\eqalign{
[\Delta_B(z), \Delta_B(w)] &=\bar L^a(z) \bar L^b(w) [L^a(z),L^b(w)] \cr
  &+ [\bar L^a(z),\bar L^b(w)] L^b(w) L^a(z)  \cr
  &+ \bar L^a(z) [L^a(z), \bar L^b(w)] L^b(w) \cr
  &+ \bar L^b(w) [\bar L^a(z),L^b(w)] L^a(z)  \cr
=& \bar L^a(z) \bar L^b(w) f^{acd} \theta^b_c(w;z) L^d(z) \cr
-& \bar L^a(z) \bar L^b(w) f^{bcd} \theta^a_c(z;w) L^d(w) \cr
+& \bar L^d(z) f^{acd} \theta^b_c(w;z) L^b(w) L^a(z)      \cr
-& \bar L^d(w) f^{bcd} \theta^a_c(z;w) L^b(w) L^a(z)      \cr
+& \bar L^a(z) f^{acd} \theta^b_c(w;z) L^d(z) L^b(w)      \cr
-& \bar L^a(z) \bar L^d(w) f^{bcd} \theta^a_c(z;w) L^b(w) \cr
+&\bar L^a(z) \Bigl\{ L^a(z) \bar L^b(w)
    +\bar L^d(w) f^{bcd} \theta^a_c(z;w)\Bigr\} L^b(w)  \cr
+& \bar L^b(w) \bar L^d(z) f^{acd} \theta^b_c(w;z) L^a(z) \cr
-& \bar L^b(w) f^{bcd} \theta^a_c(z;w) L^d(w) L^a(z)      \cr
-& \bar L^b(w) \Bigl\{ L^a(z) \bar L^b(w)
    +\bar L^d(w) f^{bcd} \theta^a_c(z;w)\Bigr\} L^a(z).  }
\eqno (E.6) $$
In the above transformations we used the commutation relation
(C.8) and the one transposed to it.
In the last expression the second term cancels the sixth one,
the fourth term cancels the ninth one. We denote the
expression in braces by $P^{ab}(z,w)$. Notice that
$$ \eqalign{
P^{ab}(z,w) &=\{ L^a(z) \bar L^b(w)
+\bar L^d(w) f^{bcd} \theta^a_c(z;w)\} \cr
&= \{ L^b(w) \bar L^a(z)
+\bar L^d(z) f^{acd} \theta^b_c(w;z)\} = P^{ba}(w,z). }
\eqno (E.7) $$

Before going further let us state a simple consequence of the
identity (C.7). Namely, setting $u \to w$ we obtain
$$\eqalign{
& L^a(z)\theta^b_c(w;w)_{reg} -L^b(w)\theta^a_c(z;w) \cr
& =-\theta^a_d(z;w)\theta^b_e(w;w)_{reg} f^{cde} \cr
& -\theta^a_d(z;w)\theta^e_c(w;w)_{reg} f^{bde}  \cr
& +\theta^b_d(w;z)\theta^e_c(z;w) f^{ade}  \cr
& +f^{bcd} \Omega^{ad}(z,w). }
\eqno (E.8) $$
This lemma allows us to rewrite $P^{ab}(z,w)$ as
$$\eqalign{
P^{ab}(z,w)=& \{-2L^b(w) L^a(z) \log\Pi - L^b(w) f^{acd}\theta^d_c(z;z) \cr
&+ f^{acd}\theta^b_c(w;z)\bar L^d(z) +L^d(z) f^{acd}\theta^b_c(w;z) \} \cr
=& \{-2L^b(w) L^a(z) \log\Pi - 2f^{acd}\theta^b_c(w;z)L^d(z)\log\Pi \cr
&- f^{acd} f^{bfe} \theta^d_f(z;w)\theta^e_c(w;z)
+2h^*\Omega^{ab}(z,w) \}.  }
\eqno (E.9) $$

We shall further use this expression, and now return to (E.6), which
after cancellations and regrouping the remaining terms becomes
$$\eqalign{
[\Delta_B(z), \Delta_B(w)] =& [\bar L^a(z), \bar L^b(w)]
f^{acd}\theta^b_c(w;z) L^d(z) -
\bar L^d(z) f^{acd} \theta^b_c(w;z) [L^a(z), L^b(w)]  \cr
&+ \bar L^a(z) P^{ab}(z,w) L^b(w)
- \bar L^b(w) P^{ab}(z,w) L^a(z)  \cr
=& L^f(z) f^{aef} \theta^b_e(w;z) f^{acd} \theta^b_c(w;z) L^d(z)  \cr
&-  L^d(z) f^{aef} \theta^b_e(w;z) f^{acd} \theta^b_c(w;z) L^f(z) \cr
&+ \bar L^d(z) \Bigl( P^{df}(z,w)+ f^{bef}\theta^a_e(z;w) f^{acd}
\theta^b_c(w;z) \Bigr) L^f(w)  \cr
&+ \bar L^f(w) \Bigl( P^{df}(z,w)+ f^{bef}\theta^a_e(z;w) f^{acd}
\theta^b_c(w;z) \Bigr) L^d(z).  }
\eqno(E.10) $$
The first two terms of the last expression cancel each other due
to the Jacobi identity (C.16).
Introduce now one more quantity $Q^{ab}(z,w)$ defined by
$$Q^{ab}(z,w)=P^{ab}(z,w)+ f^{adc} f^{bfe} \theta^d_e(z;w)
\theta^f_c(w;z).
\eqno (E.11) $$
Like for $P^{ab}(z,w)$ we have
$$ Q^{ab}(z,w)=Q^{ba}(w,z).
\eqno (E.12) $$
The expression (E.9) for $P^{ab}(z,w)$ gives
$$ Q^{ab}(z,w)=2\{ h^*\Omega^{ab}(z,w) -L^b(w) L^a(z) \log\Pi
- f^{acd} \theta^b_c(w;z) L^d(z)\log\Pi \}.
\eqno (E.13) $$
With this notation the commutator (E.10) becomes
$$\eqalign{
[\Delta_B(z), \Delta_B(w)]=&
\bar L^a(z) Q^{ab}(z,w) L^b(w) - \bar L^b(w) Q^{ab}(z,w) L^a(z)  \cr
=& \Bigl\{\bar L^a(z) Q^{ab}(z,w)\Bigr\} L^b(w)
- \Bigl\{\bar L^b(w) Q^{ab}(z,w)\Bigr\} L^a(z)
+Q^{ab}(z,w) [L^a(z), L^b(w)] \cr
=& \Bigl\{\bar L^a(z) Q^{ad}(z,w)
- Q^{ab}(z,w) f^{bcd}\theta^a_c(z;w)\Bigr\} L^d(w) \cr
&- \Bigl\{\bar L^a(w) Q^{ad}(w,z)
- Q^{ab}(w,z) f^{bcd}\theta^a_c(w;z)\Bigr\} L^d(z). }
\eqno (E.14) $$
This is a first order differential operator. It is antisymmetric
although this is not seen at once from (E.14). Therefore we can
restrict our attention to its symbol only. Returning to the
original problem (E.1), we arrive to
$$\eqalign{
[A(z), A(w)]=& Y^d(z,w) L^d(z)+\bar L^d(z) Y^d (z,w) \cr
-& Y^d(w,z) L^d(w)- \bar L^d(w) Y^d (w,z), }
\eqno (E.15) $$
where
$$ Y^d(z,w) = \Bigl\{ L^d(z) U(w)
-{1\over 2} \Bigl(\bar L^a(w) Q^{ad}(w,z)- Q^{ab}(w,z)
f^{bcd} \theta^a_c(w;z) \Bigr) \Bigr\}.
\eqno (E.16) $$
We shall prove that
$$ Y^d(z,w)=0,
\eqno (E.17) $$
then  (E.1) will follow.

Writing out $Y^d(z,w)$ explicitly,
$$\eqalign{
Y^d(z,w)=& h^*\Bigl\{ L^d(z)\Omega^{aa}(w,w)_{reg} -
\bar L^a(w) \Omega^{da}(z,w)
+ \Omega^{ab}(w,z) f^{bcd} \theta^a_c(w;z)\Bigl\} \cr
-& L^d(z)\Bigl( {1\over\Pi}\bar L^b(w) L^b(w) \Pi \Bigr) \cr
+& \bar L^a(w)L^d(z)L^a(w)\log\Pi  \cr
+& \bar L^a(w) f^{acb} \theta^d_c(z;w) L^b(w) \log\Pi \cr
-& f^{bcd}\theta^a_c(w;z) L^b(z) L^a(w) \log\Pi \cr
-& f^{bcd}\theta^a_c(w;z) f^{aef} \theta^b_e(z;w) L^f(w) \log\Pi.  }
\eqno(E.18) $$
Transforming the expression in braces, we arrive to
$$\eqalign{
&\Bigl\{ L^d(z)\Omega^{aa}(w,w)_{reg} -
\bar L^a(w) \Omega^{da}(z,w)
+ \Omega^{ab}(w,z) f^{bcd} \theta^a_c(w;z)\Bigl\} \cr
&\qquad =\Bigl\{ -\Omega^{da}(z,w) \bar L^a(w) + \Omega^{ba}(z,w)
f^{bcd}\theta^a_c(w;z) \cr
&\qquad \qquad+\Omega^{ea}(z,w)\theta^a_f(w;z) f^{dfe}
+\Omega^{df}(z,w)\Xi^f(w) \Bigr\} \cr
&\qquad =\Bigl\{ 2\Omega^{df}(z,w) L^f(w) \log\Pi\Bigr\}. }
\eqno(E.19) $$
Two next summands in (E.18) become
$$\eqalign{
 \bar L^a(w)L^d(z)L^a(w)\log\Pi
-& L^d(z)\Bigl( {1\over\Pi}\bar L^b(w) L^b(w) \Pi \Bigr) \cr
=[\bar L^a(w), L^d(z)] L^a(w) \log\Pi -& 2\Bigl\{L^a(w) \log\Pi\Bigr\}
 L^d(z) L^a(w)\log\Pi \cr
=& \bar L^b(w) f^{acb}\theta^d_c(z;w) L^a(w) \log\Pi \cr
-& f^{dcb}\theta^a_c(w;z) L^b(z) L^a(w)\log\Pi \cr
-& \{L^d(z) \bar L^a(w) + \bar L^b(w) f^{acb}\theta^d_c(z;w)\}
L^a(w)\log\Pi \cr
-& \{ 2L^d(z) L^a(w)\log\Pi\} L^a(w)\log\Pi \cr
&\qquad = \bar L^b(w) f^{acb}\theta^d_c(z;w) L^a(w) \log\Pi \cr
&\qquad - f^{dcb}\theta^a_c(w;z) L^b(z) L^a(w)\log\Pi \cr
&\qquad -\Bigl\{ L^d(z)\Xi^a(w) + \bar L^b(w) f^{acb}
\theta^d_c(z;w) \Bigr\} L^a(w)\log\Pi. }
\eqno(E.20) $$
The first two terms cancel the fourth and the fifth summands in
(E.18) thus leading to
$$\eqalign{
Y^d(z,w)=& 2h^*\{\Omega^{df}(z,w) L^f(w) \log\Pi\} \cr
&\qquad - \Bigl\{ f^{efa} f^{cbd} \theta^f_c(w;z) \theta^b_e(z;w) \cr
&\qquad\qquad +L^d(z)\Xi^a(w)+ \bar L^b(w) f^{acb} \theta^d_c(z;w) \Bigr\}
L^a(w) \log\Pi. }
\eqno(E.21) $$
Notice that
$$\eqalign{
& L^d(z)\Xi^a(w)+ \bar L^b(w) f^{acb} \theta^d_c(z;w)  \cr
&\qquad =f^{acb}\Bigl( L^b(w)\theta^d_c(z;w)
- L^d(z)\theta^b_c(w;w)_{reg} \Bigr)  \cr
&\qquad\qquad +\theta^d_c(z;w) f^{cba} \Bigl( \Xi^b(w)
-2 L^b(w)\log\Pi \Bigr) . }
\eqno (E.22) $$
The very last term $\{ 2\theta^d_c(z;w) f^{cba} L^b(w)\log\Pi \} $
vanishes after multiplying by $L^a(w)\log\Pi$. The first
term in (E.22) can be rewritten according to (E.8), which
leads to
$$\eqalign{
Y^d(z,w) &= 2h^*\Omega^{df}(z,w) L^f(w)\log\Pi \cr
&-\Bigl\{ f^{efa} f^{cbd} \theta^f_c(w;z) \theta^b_e(z;w) \cr
&\qquad +\theta^d_c(z;w) f^{cba}\Xi^b(w)  \cr
&\qquad +f^{acb} \Bigl(\theta^d_f(z;w)\theta^b_e(w;w)_{reg} f^{cfe} \cr
&\qquad\qquad +\theta^d_f(z;w)\theta^e_c(w;w)_{reg} f^{bfe}   \cr
&\qquad\qquad -\theta^b_f(w;z)\theta^e_c(z;w) f^{dfe}  \cr
&\qquad\qquad -f^{bcf}\Omega^{df}(z,w)\Bigr) \Bigr\}
L^a(w)\log\Pi   \cr
&\qquad\qquad\qquad =0. }
\eqno (E.23) $$
Here the first line cancels the last one; the second line --- the
sixth one; the third to fifth lines cancel due to the
Jacobi identity.

\filbreak
\bigskip
\centerline{References.}
\bigskip
\nobreak
\item{[1]} E.Witten, Comm. Math. Phys. 92 (1984) 455.
\item{[2]} V.G.Knizhnik and A.B.Zamolodchikov, Nucl. Phys. B247 (1984) 83.
\item{[3]} D.Bernard, Nucl. Phys. B303 (1988) 77.
\item{[4]} D.Bernard, Nucl. Phys. B309 (1988) 145.
\item{[5]} A.Losev, {\it Coset construction and
Bernard equations}, preprint CERN-TH.6215/91.
\item{[6]} E.Martinec, Nucl. Phys. B281 (1987) 157.
\item{[7]} O.Alvarez and P.Windey, {\it The energy-momentum
tensor as geometrical datum}, preprint UCB-PTH-86/36.
\item{[8]} S.Axelrod, S.Della Pietra and E.Witten, J. Diff. Geom.
33 (1991) 787.

\end